# Insights into the enhancement of oxygen mass transport properties of strontium-doped lanthanum manganite interface-dominated thin films.

F. Chiabrera[1], A. Morata[1], M. Pacios[1], A. Tarancón[1,*]

[1]*Department of Advanced Materials for Energy, Catalonia Institute for Energy Research (IREC), Jardins de les Dones de Negre 1, Planta 2, E-08930 Sant Adrià de Besòs (Barcelona), Spain, e-mail: atarancon@irec.cat*

## Abstract

Strontium-doped lanthanum manganite thin films were deposited by pulsed laser deposition on yttria-stabilized zirconia single crystals for a comprehensive electrochemical characterization of the material acting as a cathode. A physically-meaningful electrical model was employed to fit the electrochemical impedance spectroscopy results in order to extract the main oxygen mass transport parameters as a function of the temperature and oxygen partial pressure. The oxygen diffusion and surface exchange coefficients extracted from the analysis showed several orders of magnitude of enhancement with respect to the bulk values reported in the literature and an unexpectedly low dependence with the oxygen partial pressure. Different observations were combined to propose a mechanism for the enhanced incorporation of oxygen in interface-dominated thin films mainly based on the high concentration of oxygen vacancies expected in the grain boundaries.

**Keywords:** LSM, grain boundary, oxygen mass transport, thin film, MIEC

# 1. Introduction

Sr-doped lanthanum manganite (LSM) is one of the most studied materials for solid oxide fuel cell cathodes, which was chosen among others for its high electronic conductivity, good high temperature mechanical compatibility with Yttria-Stabilized Zirconia (YSZ) and low fabrication cost [1–6]. On the other hand, LSM is known to be a poor oxygen conductor, leading to a cathode active surface that, although it has been found not to be confined to the triple phase boundary air/electrolyte/LSM, is very limited, causing high polarization losses [2,5,6]. For this reason, many efforts have been devoted to the study of Mixed Ionic-Electronic Conductors (MIEC), eventually able to extend the chemical active zone to the entire cathode volume, and to the understanding of the main mechanisms involved in the complex behavior of oxygen reduction and transport [7,8]. Beyond the research of unusual materials with superior oxygen transport and reduction properties, which can reveal commercialization issues [9], many studies in the last few years have focused on the improvement of well-known composites, by tuning the interaction between structural and electrochemical properties [10,11]. Recent independent works demonstrated that dense LSM thin films with columnar nanometric grains present several orders of magnitude of increase of oxygen diffusion and surface exchange coefficient [12,13]. Although the exact mechanism is still unknown, the large enhancement seems to come from the grain boundaries, where a high concentration of strain-induced defects tends to accumulate.

In this work, the electrochemical transport properties of interface-dominated columnar nanometric $La_{0.8}Sr_{0.2}MnO_{3+\delta}$ thin films are studied as a function of oxygen partial pressure and temperature by means of Electrochemical Impedance Spectroscopy (EIS). The dependence of the main mass transport parameters is derived in order to discriminate the differences between the grain boundary- and bulk-governed LSM behaviors and proposing a mechanism for the oxygen incorporation able to explain the enhancement of the oxygen reduction reaction (ORR) performance observed in thin films.

## 2. Methodology

LSM thin films of 100 nm (LSM100) and 200 nm (LSM200) were deposited on (100)-oriented YSZ single crystals by Pulsed Laser Deposition (PLD) in a large-area system from PVD Products (PLD-5000) with a KrF-248 nm excimer laser from Lambda Physics (COMPex PRO 205). The films were deposited with an energy fluency of 1 J cm$^{-2}$ per pulse at a frequency of 10 Hz. The substrate temperature of deposition was held at 700ºC, the oxygen partial pressure at 0.026 bar and the target-to-substrate distance was fixed at 95 mm. After deposition, all the samples were subjected to an annealing process in air at 700ºC.

The LSM films microstructure was analyzed by Scanning electron microscope (SEM) in a ZEISS AURIGA equipment. The surface of the films was studied by Atomic Force microscope (AFM, XE 100 Park System Corp.) in intermittent mode. The images were analyzed in the XEI software (Park System Corp.) to extract the root-mean square roughness (RMS) and the grain boundary length per unit area. Structural studies were conducted by X-Ray diffraction (XRD) in a Bruker D8 Advance diffractometer system. A Bragg–Brentano theta-2 theta configuration was used, applying an offset to reduce the contribution from the single crystalline substrate.

Electrochemical impedance spectroscopy (EIS) was performed on a symmetrical LSM/YSZ$_{SC}$/LSM cell with a Novocontrol Alpha-A analyzer. The frequency range chosen was $10^6$-0.05 Hz and an ac voltage with amplitude 0.05 V was applied. The experiments were carried out inside a ProboStat test station (NorECs) placed inside a vertical furnace. The oxygen partial pressure was controlled with a ZIROX oxygen pump and confirmed by a lambda-sensor located next to the analyzed sample. The temperature was varied from 600ºC to 700ºC after a stabilization at the lower oxygen partial pressure for 12 hours. The total area of the symmetrical cell was measured to be 0.64 cm$^2$. In order to improve the current collection on the LSM thin film, a porous layer of gold was painted on both sides of the cell.

Distribution of Relaxation Time (DRT) analysis was conducted to derive preliminary insights in the nature of the processes involved in the collected impedance spectra. DRT is a powerful tool to distinguish the different phenomena that contribute to the impedance spectrum and to calculate the resistive contribution of each of them [14]. It is based on the assumption that every impedance that can be described by the Kramers-Kronig relations, therefore attributable to a physical response, can be also divided into a series of infinitesimal RC elements, each one characterized by a characteristic time $\tau_i=1/R_iC_i$. Because of the mathematical complexity of the problem, the deconvolution of the measured impedance into the characteristic relaxation time is challenging. Among the various approaches that have been proposed in the literature, such as digital filtering of Fourier transform [15] or Genetic Programming [16], the Radial Basis Function DRT recently developed by Ciucci and coworkers [17] has been chosen. In their work, they developed a MATLAB GUI toolbox that allows using different basis functions, such as piecewise linear or Gaussian, to perform DRT analysis of the measured data. In this work, Gaussian based function was chosen and an analysis on each spectrum collected was conducted to find the best regularization parameters.

## 3. Results and Discussion

### 3.1 Microstructural Characterization

In this section, the structural and microstructural characterization of the deposited LSM thin films is briefly presented (a comprehensive characterization can be found elsewhere for identically deposited samples [12]). Figure 1a and 1b show the AFM surface of the 100 nm and 200 nm film respectively. The samples present both a very high quality of the surface, with nanometric grains and a calculated RMS is 0.44 nm for the LSM100 and 0.59 nm for the LSM200. Despite the small difference in surface roughness, LSM200 presents a more heterogeneous surface than the LSM100, as highlighted by the larger variance of grain sizes. The exposed grain boundary length per unit area is estimated to be 88 µm/ µm$^2$ for the LSM100 and 75 µm/ µm$^2$ for the LSM200 (Figure S1, supplementary info). The values calculated are similar to those reported by Usiskin *et al.* for the samples deposited at low temperature (below 600ºC) while strongly differ if confronted to the same temperature of deposition (700ºC) [18]. This is probably due to the different stoichiometry of the LSM target, a A-site deficient $(La_{0.8}Sr_{0.2})_{0.95}MnO_{3+\delta}$ was used in their work, as already highlighted by Navickas *et al.* [13]. Figure 1c show a cross section SEM picture of a representative LSM layer deposited on an YSZ single crystal after annealing at 700ºC. A fully dense columnar structure with nanometric grain size is observed. This dense vertically aligned microstructure is typically obtained for thin films fabricated with PLD when high temperature and low pO$_2$ conditions are employed [12]. Figure 1d shows the XRD diffractogram obtained from the same sample. The data was indexed to a rhombohedral *R3c* space group (*SG #167*) corresponding to LSM. No further peaks from eventual parasitic phases are detected. Differences found in the relative intensity of the peaks compared with the indicated by the pattern (JCPDS 00-051–0409) for the different reflections indicate a certain preferential orientation of the crystallites. Single phase dense and homogeneous polycrystalline thin films of LSM were deposited on both sides of YSZ single crystals for electrochemical characterization.

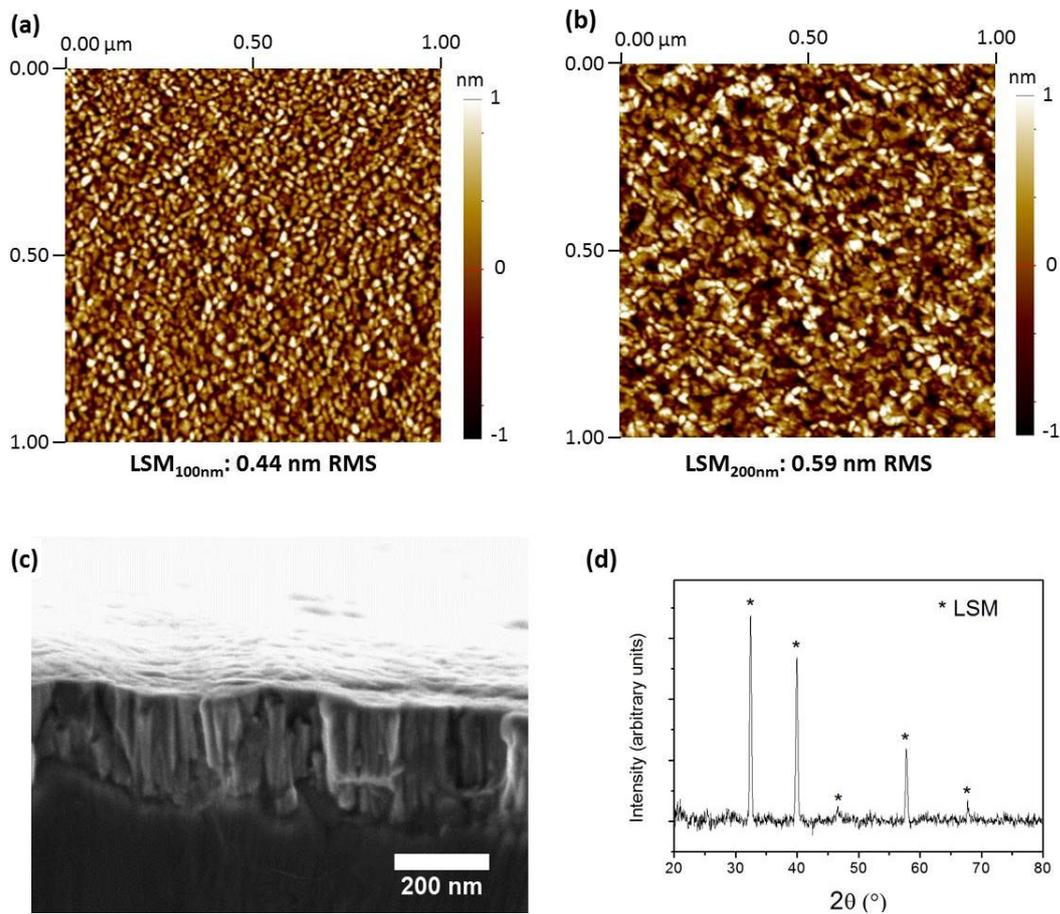

**Figure 1.** Surface AFM image of a 100 nm (a) and a 200nm (b) PLD-deposited LSM thin film on YSZ single crystal. (c) Cross-section of a 200nm LSM film on YSZ single crystal showing the characteristic columnar structure (d) Typical XRD diffraction patterns of the LSM layers deposited on YSZ single crystals (an offset was applied to remove the contribution from the substrate).

## 3.2 Electrochemical characterization

Dense LSM/YSZ/LSM symmetrical cells were characterized by electrochemical impedance spectroscopy under different temperature and $pO_2$ conditions. The Nyquist plot obtained for the 100 nm and 200 nm cell at 700ºC in synthetic air is shown in figure 2a. Both spectra are characterized by two major contributions corresponding to low and high frequencies (visible in the inset of figure 2a). The distributions of relaxation times associated to these EIS spectra (figure 2b) show a series of smaller peaks at low characteristic times ($P_{D1}$, $P_{D2}$) together with a larger contribution to the total impedance at $\tau \approx 0.01s$ ($P_s$). The series of $P_{Di}$ peaks with increasing intensity with time can be ascribed to a diffusion process [14,19,20] such as the oxygen diffusion in LSM thin films. On the other hand, the characteristic time and shape (ZARC element) of the other contribution indicates that this impedance is associated to the oxygen incorporation reactions at the surface of the LSM layer. Accordingly, the diffusion and surface reaction processes are clearly distinguishable being the surface resistance the responsible of the main contribution to the total impedance for all the samples analyzed. Confronting the spectra of the two samples in Figure 2a, it is possible to notice that the 200 nm sample show a larger resistance of both diffusion and ORR processes respect to the 100 nm LSM film (also shown in the total polarization resistance on temperature in Fig. S2, supplementary info). The higher diffusion resistance is emphasized in the DRT analysis by an increase and a shift of the $P_{D1}$, $P_{D2}$ peaks, as expected by the increase of sample thickness (Fig. 2b). The change in surface reaction resistance can be ascribed to the higher concentration of grain boundary in the surface of the 100 nm sample respect to the 200 nm one (Fig. S1, supplementary info). As reported by Usiskin *et al.*, the ORR resistance in LSM thin films shows a strong dependence on the surface grain boundary density, determining the best performance (i.e. lower resistance) for the sample with more exposed grain boundaries [18].

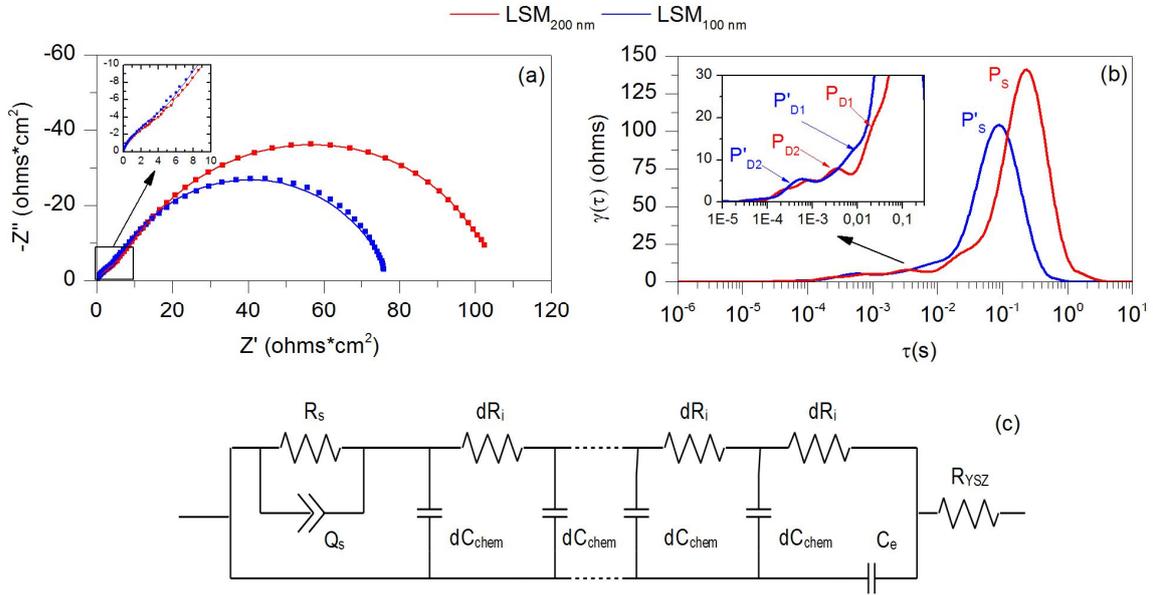

**Figure 2.** Impedance spectra measured on 100 nm and 200nm LSM symmetrical thin films on YSZ single crystal at 650ºC in synthetic air (in blue and red respectively). (a) Nyquist plot of the EIS measurements (squares) and corresponding fitted model (solid line); (b) DRT plot of the EIS spectra contained in (a); (c) equivalent circuit employed to fit the measured impedance spectra.

In order to reproduce the EIS behavior of a dense thin film of a MIEC deposited on a pure ionic conductor electrolyte, the equivalent circuit derived by Jamnik and Maier [21] (and readapted in the case of LSM thin films by Fleig and co-workers [3]) was employed for this work. Figure 2c shows the equivalent circuit used here to fit the impedance spectra. It includes: a resistance and CPE in parallel ($R_s$-$Q_s$), which accounts for the surface oxygen incorporation processes, a transmission line with infinitesimal resistances ($dR_i$) and capacitance ($dC_{chem}$) in parallel that reproduces the solution of the diffusion equation [22] and a pure capacitance ($C_e$) describing the blocking electrons surface between LSM and YSZ. At the end of the circuit, a resistance in series ($R_{YSZ}$) is placed, which corresponds to the pure resistive behavior of the YSZ single crystal at high temperatures. The adopted circuit differs from the model of Fleig and co-workers [3] in the position of the CPE, here shifted in parallel with the surface resistance. This solution has been already adopted in a similar work to enhance the effectiveness of the fitting [18] and it is particularly necessary at low oxygen partial pressure, while in air different equivalent circuits can be adopted [12]. Figure 2b shows the high quality of a typical fitting of the measured impedance spectra by using this model. The accurate fitting means that the electrochemical oxygen reduction reactions (ORR) take place mostly at the LSM surface, followed in series by oxygen diffusion in the cathode thin film in order to reach the YSZ electrolyte.

The parameters obtained in the EIS can be used to extract the overall surface exchange coefficient $k^q_{av}$ and the oxygen diffusion $D^q_{av}$. In this study, these mass transport parameters are an average between the grain and grain boundary value and represent the overall properties of the LSM thin film measurable by EIS. As derived from the Nerst-Einstein equation, the diffusion coefficient is calculated by:

$$D^q_{av} = \frac{L}{R_{ion} \cdot A} \cdot \frac{k_b T}{c_{O2} z_i^2 e^2} = \sigma_{ion} \cdot \frac{k_b T}{c_{O2} z_i^2 e^2} \quad (1)$$

Where $L$ and $A$ are respectively the thickness and the surface area of the LSM, $k_b$ is the Boltzmann constant, $z_i$ is the number of charge involved in the transport, $e$ is the elementary charge and $c_{O2}$ is the concentration of oxygen in the LSM lattice. The coefficient $D^q$ can be confronted with the diffusion coefficient $D^*$ obtained in the oxygen tracer experiments considering a correlation factor of $f$=0.69 [23,24]. The surface exchange coefficient is given by:

$$k_{av}^q = \frac{1}{R_s \cdot A} \cdot \frac{k_b T}{z_i^2 e^2 c_{O_2}}$$  (2)

The confrontation between $k^q$ and $k^*$ (being $k^*$ the oxygen self-diffusion exchange coefficient) is still possible, although some differences could rise from the use of a catalytic active current collector [23]. For this reason, gold paste and gold cables were used in the measurements, being gold known to be mostly inactive for ORR reactions [25]. Since in LSM thin films with columnar nanometric grain size the grain boundaries are believed to offer a direct pathway for the incorporation and the diffusion of oxygen (a very low contribution is expected from the grain bulk [12,13,18] ), the adsorption sites inactive for the ORR will be also inactive for the oxygen diffusion. On these bases, the ratio k/D is not expected to vary for the presence of the porous gold paste.

### 3.2.1 Oxygen mass transport properties as a function of temperature

Impedance spectra measured at different temperatures in air are shown in figure 3. The characteristic response of a MIEC-based symmetrical cell (figure 2a) is observed in all the spectra. The Nyquist plots in figure 3a and 3b, as well as the Bode plot in figure 3c, shows that the contributions at high frequencies, attributable to oxygen diffusion, and at low frequencies, ascribable to surface reactions, increase their resistance when decreasing the temperature. This behavior is confirmed in the DRT plot (figure 3d) where is clearly visible an increase of the low frequency peak, along with a characteristic expansion of the peaks typically observed for a diffusion process.

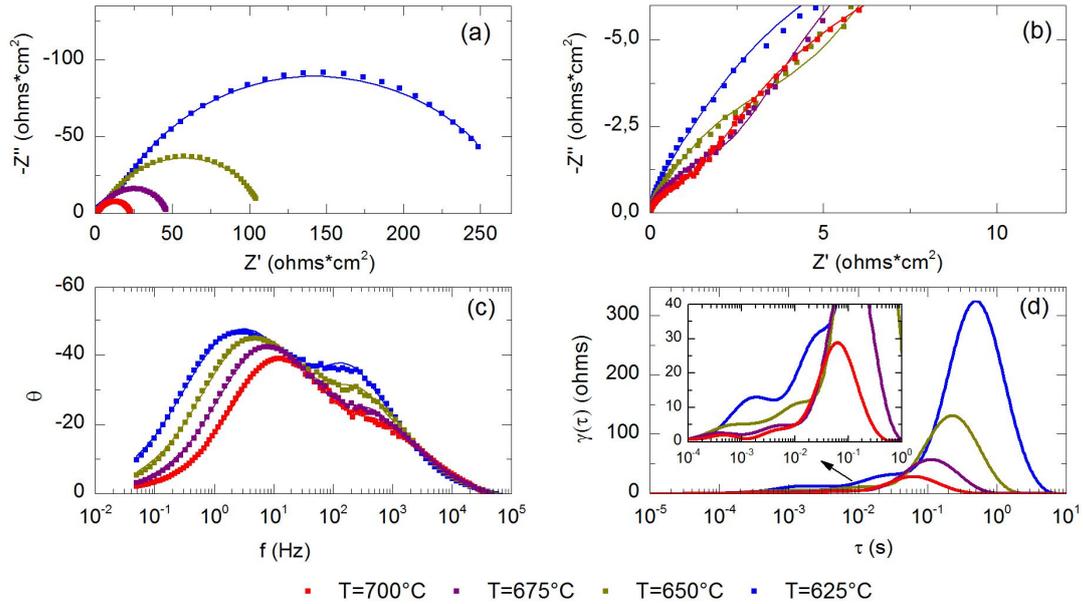

**Figure 3.** Impedance spectra for 200 nm LSM/YSZ/LSM symmetrical cells measured at different temperatures in synthetic air. (a) Nyquist plot and (b) zoom at high frequencies; (c) Bode plot; (d) DRT plot for different temperatures. All the contributions increase their resistance when decreasing the temperature.

Figure 4a shows the evolution with the temperature of the oxygen diffusion ($D^q_{av}$) calculated from the fitted impedance spectra (two samples with different thickness were analyzed showing similar results). An Arrhenius behavior can be observed for the oxygen diffusion with an activation energy of $E_a$=2.6 eV. For comparison, Figure 4a also includes some diffusion values previously reported in the literature for LSM. All these values are derived from LSM thin films deposited by PLD, except for the reference values measured by De Souza and co-workers for LSM bulk diffusion [2,26]. Grain ($D_g^*$) and grain boundary ($D_{gb}^*$) oxygen self-diffusion coefficients measured using Isotope Exchange Depth Profiling coupled to Secondary Ion Mass Spectroscopy (IEDP-SIMS) by Navickas *et al.* [13] and by Saranya *et al.* [12] were averaged in this plot to make them comparable with other results coming from EIS. This is necessary because the electrochemical analysis is unable to distinguish grain and grain boundary returning a response corresponding to the convolution of both contributions. By comparing the averaged coefficients, a collection of values within the range of two orders of magnitude is observed illustrating a great scattering of results. The diffusion measured in this work by EIS shows the highest values with an activation energy similar to the bulk values reported in the reference work by De Souza *et al.* ($E_a$=2.8 eV) [2]. However, similarly derived diffusion coefficients for PLD-deposited LSM layers by Navickas *et al.* [13] show values two orders of magnitude below at 700ºC. Averaged grain and grain boundary values obtained after IEDP-SIMS experiments remain in between both EIS studies. Although IEDP-SIMS measurements are typically considered accurate direct measurements of the self-diffusion coefficients, in this case, strong assumptions were necessary to adjust the diffusion profiles (since simplistic numerical models were employed). For this reason, the arising values for D* and k* might be taken as simple approximations.

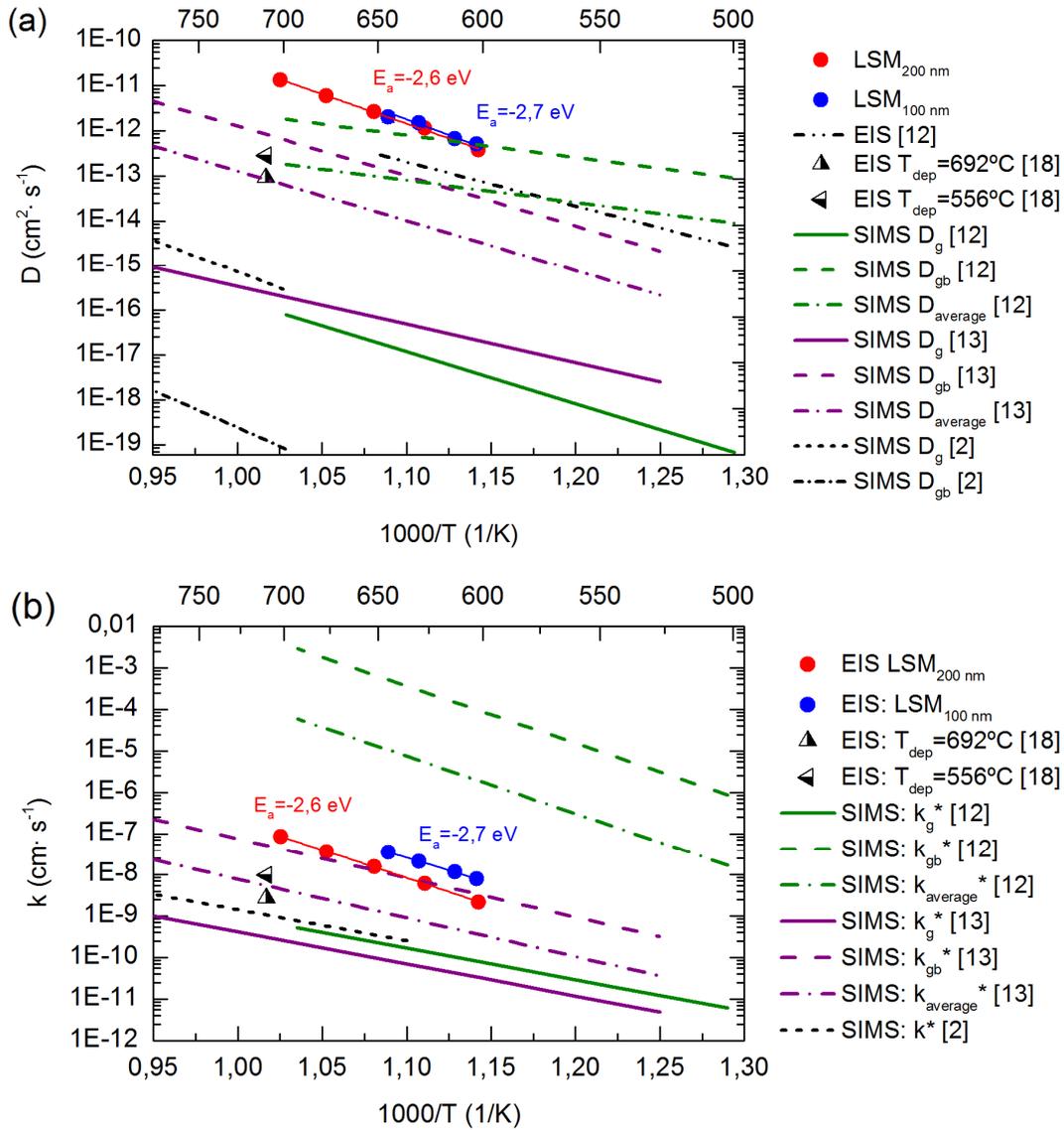

**Figure 4.** Temperature dependence of the ionic diffusion coefficient (a) and the surface exchange coefficient (b) calculated in air for the 200 nm and the 100 nm sample. In the plot are included D* and k* derived from [12] (green lines: solid bulk $D^*_b$ and $k^*_b$, dashed grain boundary $D^*_{gb}$ and $k^*_{gb}$, dash dotted averaged $D^*_{av}$ and $k^*_{av}$; black dash dot-dot line for $D^q_{av}$ obtained by EIS for a 30 nm LSM thin film); from [13] (purple lines, solid grain $D^*_b$ and $k^*_b$, dashed grain boundary $D^*_{gb}$ and $k^*_{gb}$, dash dotted averaged $D^*_{av}$ and $k^*_{av}$); from [2] (black dashed for bulk $D^*_b$ and $k^*_b$, black short dash dotted for grain boundary $D^*_{gb}$) and from [18] (triangle for $D^q$ and $k^q$ obtained by EIS

for samples presenting high ($T_{dep}$=556°C) and low ($T_{dep}$=692°C) grain boundary coverage).

Figure 4b shows the calculated oxygen surface exchange coefficient ($k^q_{av}$) evolution with temperature. Remarkably, the activation energy has been calculated to be $E_a$=2.6 eV, exactly the same as obtained for diffusion, which could indicate a close relationship between the two processes. An activation energy above 2 eV suggests that the bulk pathway is the predominant (for triple phase boundary pathways an activation energy near 1.5 eV is expected [27]). Figure 4b also shows a comparison with the values of $k$ reported in the works previously included in figure 4a. Similar to the diffusion, a great scatter of results is observed for the oxygen surface exchange coefficients. The values obtained in this work are one order of magnitude higher than averaged values reported in references [13] and [18] while two orders of magnitude lower than those from Saranya *et al.* [12].

Although all these LSM thin films were deposited by PLD, significantly higher values of oxygen diffusion and surface exchange coefficients are systematically obtained by Tarancón and co-workers using EIS (this work) and IEDP-SIMS [12]. This is probably due to major differences on microstructural, compositional and/or state of strain of the LSM layers derived from the use of different deposition conditions and substrates. Excluding important deposition parameters such as temperature, oxygen partial pressure and laser fluency (Usiskin *et al.* studied libraries of microelectrodes deposited under a large variety of conditions in reference [18]), the target-to-substrate distance could significantly influence the quality of the deposited thin film. In this direction, long target-to-substrate distances are typically employed by large-area PLD deposition systems used by Tarancón and co-workers while much shorter ones are reported by Navickas *et al* [13]and Usiskin *et al* [18]. Additionally, an important effect can also be derived from the use of different substrates. Silicon-supported substrates (employed by Tarancón and co-workers for IEDP-SIMS in reference [12]) will generate a higher strain than YSZ substrates due to a higher thermal mismatch between Silicon and YSZ/LSM. This can increase the defect concentration at the grain boundary level and the total strain in the thin film, which has been demonstrated to affect its transport properties [28,29]. Other differences can rise from the measurement methodology. In particular, the presence of a metallic current collector (this work) can enhance the homogeneity of current distribution along the surface ensuring no ORR inactive zone, while probing directly the thin film can create additional losses due to the electronic sheet resistance [30]. In any case, it is clear that further work is still needed to determine the origin of the remarkable differences observed between samples in order to understand and fully control the oxygen mass transport enhancement reported in nanostructured thin films.

**3.2.2 Oxygen mass transport properties as a function of the oxygen partial pressure**

Figure 5 shows the evolution of the electrochemical impedance for three selected oxygen partial pressures ranging from $10^{-1}$ to $10^{-5}$ bar at 690ºC. The part of the impedance spectrum associated with the surface resistance shows a clear increase of the polarization by decreasing the atmosphere oxygen content, as shown in the Nyquist plot of figure 5a. The magnification of the high frequency region (figure 5b) reveals a small evolution of the contribution associated to the oxygen diffusion, which appears to change in shape (see fig. 5c). This is confirmed by the DRT analysis (fig. 5d). A large increase of the peak associated to the surface resistance ($P_s$) is shown but just a small change in the diffusion peaks ($P_{Di}$) takes place. It is interesting to notice how, decreasing the oxygen content, the diffusion peaks become sharper and tend to separate from the low frequency contribution. A shift of the characteristic frequency in the DRT plot is related with a change in the capacitance or in the resistance associated with that specific process [15]. In the case of the low frequency peak $P_s$ the shift is clearly associated with an increase of the resistance. The diffusion peaks show a small increase of the maximum resistance while the area underneath them (associated with the total resistance of the process) remains almost constant. Hence, the shift of the characteristic frequency is probably related to an increase of the related capacitance, i.e. the chemical capacitance.

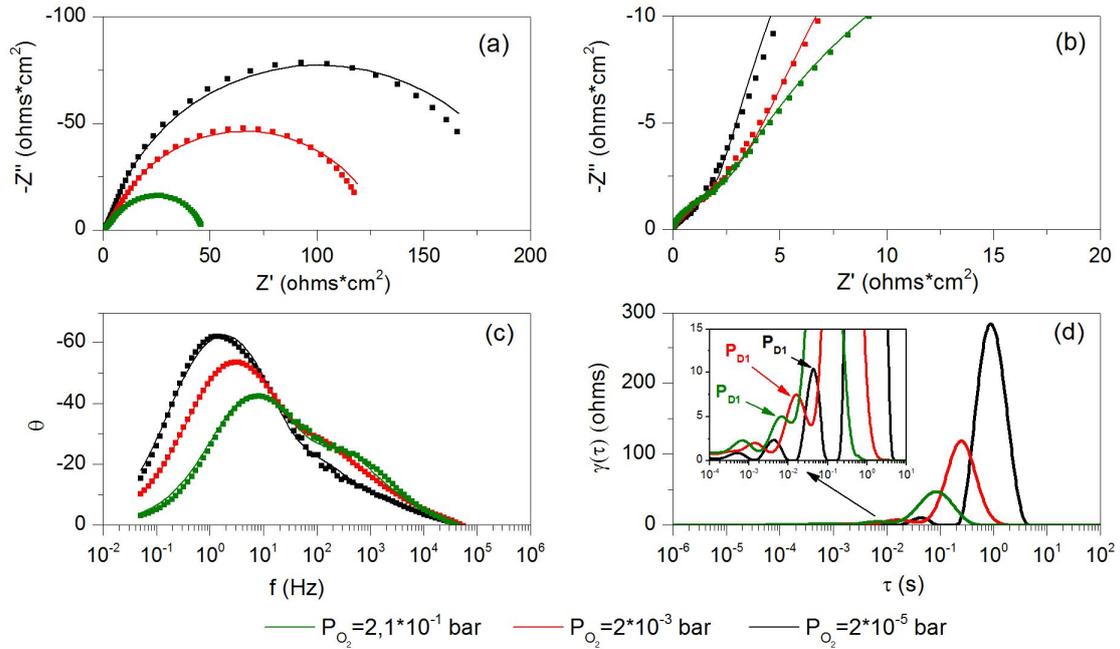

**Figure 5.** Impedance spectra measured in the 200 nm sample at different oxygen partial pressure at 675ºC. The Nyquist plot is shown in (a) and in the enlargement (b), while the Bode plot is shown in (c). In (d) is reported the DRT of the reported spectra. It is possible to notice an increase of the surface resistive contribution while the resistive contribution of the diffusion remains almost constant.

In order to quantitatively extract the dependence of the main transport parameters, the fitting of the impedance spectra obtained at different $P_{O2}$ is carried out by employing the model presented in figure 2c. Figure 6a shows the change of oxygen diffusion for the LSM thin films as a function of the oxygen partial pressure at 650ºC and 700ºC. For comparison, Figure 6a also includes some results extracted from literature for bulk [3] and similar LSM layers [18]. As previously pointed out in section 3.2.1, diffusion values several orders of magnitude higher than other works have been obtained in this study, also for different partial pressures. In contrast with the typical strong dependences expected for LSM [3], the diffusion shows a small dependence on pO$_2$ with a calculated slope of the log-log plot *m=*-0.05 in the entire set of measurements. A similar unexpected behavior was also reported by Usiskin *et al* [18], particularly for LSM thin film layers deposited by PLD at high temperature.

In general, for diluted systems is expected an oxygen diffusivity proportional to the oxygen vacancy concentration [23]. Therefore, a change in the $P_{O2}$ should modify the defect chemistry of the oxide, eventually leading to a large variation of the oxygen stoichiometry (depending on the defect chemistry of the material) [31]. In particular, hyper-stoichiometric perovskite materials with very low values of diffusion in air conditions, such as LSM [4], increase their conductivity by decreasing the $P_{O2}$ due to an increase of the oxygen vacancies concentration [2,3]. The low dependence found in this work indicates that a high concentration of oxygen vacancies is already present in air (as suggested in [12]) and that the change in $P_{O2}$ is not able to modify the equilibrium. This behavior could be explained by a formation of defect clusters in the grain boundaries, creating an electro-chemo-mechanical field that locally modifies the defect chemistry of the LSM thin film. Accordingly, a change in partial pressure (or a change in the bias applied, as found in [32]) could influence the grain bulk but just slightly the grain boundary even for very large $P_{O2}$ variations.

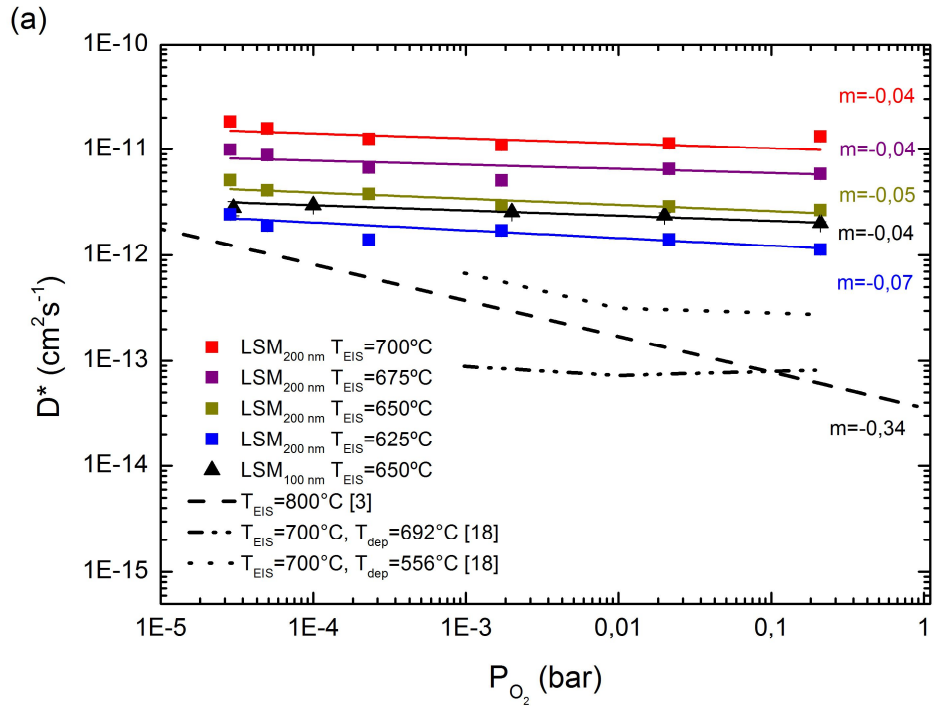

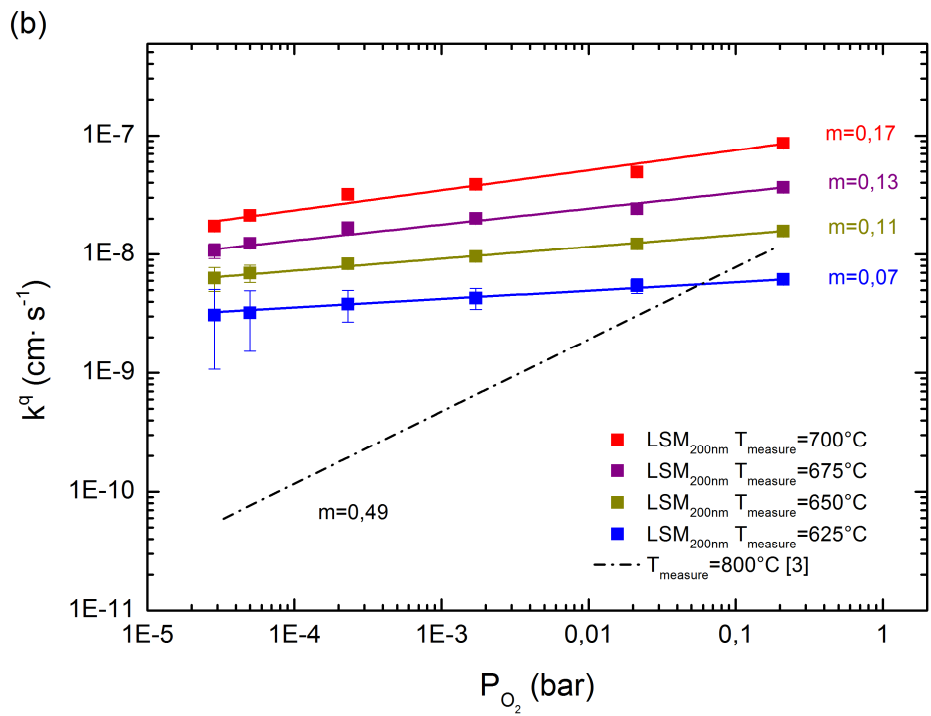

**Figure 6.** (a) Partial pressure dependence of the ionic diffusion of the 200 nm-thick sample at 625°C, 650°C, 675°C and 700°C and of the 100nm-thick sample at 650°C. Diffusion coefficient extrapolated from the conductivity of LSM obtained in [3] (dashed line) and in [18], for samples with low and high grain boundary coverage (dashed dot-dot line and short dashed respectively), are included. (b) Partial pressure dependence of the surface exchange coefficient of the 200 nm-thick sample at 625°C, 650°C, 675°C and 700°C. Surface exchange extrapolated from [3] (dashed line) is included.

Information of the defect chemistry of a material can be extracted from the chemical capacitance variation with the oxygen content and the temperature [33]. The chemical capacitance is a measure of the chemical energy stored in a material under the application of an electrical bias [21]. As noticed in the DRT analysis, the volumetric chemical capacitance ($\varepsilon_{chem}=C_{chem}/Volume_{(film)}$) is observed to increase when reducing the $P_{O2}$, with a slope in the log-log plot of m≈0.2 (figure S3, supplementary info), while in other works a more constant behavior has been measured [3]. It is surprising to notice that, despite the large difference observed in the main transport parameters, $\varepsilon_{chem}$ of different works is similar. The reason could be that, although the oxygen transport takes place mostly in the grain boundaries, the chemical response of the bulk is still accessible to the EIS measurements. Indeed, since the electronic pathway extends also in the grain (parallel to the diffusion in the equivalent circuit adopted, Fig 2c), the chemical capacitance will correspond to the response of the bulk to the chemical potential gradient induced by the diffusion in the grain boundary. Unfortunately, due to the complex defect chemistry of LSM [34] it is not trivial to extract chemical information from the change of the $C_{chem}$.

Figure 6b describes the $P_{O2}$ dependence of the oxygen surface exchange coefficient $k^q_{av}$

measured by EIS at different temperatures. The data exhibits a slope $m$ ($k^q_{av} \propto P_{O2}^m$) equal to 0.17 at 700°C. By lowering the temperature, the slope continuously decreases. There is a strong discrepancy of these results with the slope values of $m=0.49$ usually found in the literature for the bulk material at 800°C [3]. In principle, the evolution of $k^q$ with the $P_{O2}$ is strictly correlated with the nature of the oxygen involved in the rate-determining step (*rds*) of the oxygen incorporation on the MIEC surface. According to the literature, values of $m$ higher than 0.5 are related to one or more molecular species (such as $O_2^-$ or $O_2^{2-}$) while a slope of 0.5 is found for atomic species controlling the *rds* ($O^-$) [35]. High vacancy and electron concentrations contribute to lower the slope, due to their general negative dependence [36], leading to a value of $m \approx 0.25$ for a *rds* based on charge transfer mechanisms [25,37]. Even lower values of slope are ascribable to a nonlinear correlation between the oxygen surface coverage and the $PO_2$, since charged species at the surface can generate a potential step that decreases the partial pressure dependence of $k^q$ [38]. This behavior is more evident at low temperatures, due to the higher oxygen chemisorption barrier. Therefore, the hereby observed unusually low $P_{O2}$ dependence for LSM surface exchange coefficient is in good agreement with the possibility of a rate determining step involving monoatomic oxygen species ($O^-$).

Additional interesting observations for complementing the picture of the main probable ORR mechanisms arise by studying the activation energy of $k^q$ at different oxygen partial pressures (Figure S4, supplementary info). It has been found that decreasing the $P_{O2}$, the surface exchange decreases its activation energy, from 2.65 eV in air to 1.59 eV at ~$10^{-5}$ bar. This evident change involves a modification of the ORR rate-limiting step. Maier and co-workers studied the most favorable pathways for oxygen incorporation in LSM [39]. They found that for high $P_{O2}$ the approach of a vacancy to an absorbed $O^-$ is the rate limiting step (overall activation energy expected 4.0-2.7eV and $P_{O2}$ dependence *m*<0.5), while for low $P_{O2}$ dissociation of $O_2^{j-}$ without vacancy involved becomes the *rds* (overall activation energy expected 0.6-0.5 eV and $P_{O2}$ dependence m=0 - 1). The switch from one to the other rds is due to the variation of molecular adsorbate coverage and oxygen vacancy concentration that takes place in bulk LSM for a change in $P_{O2}$. The energy of activation and the scarce partial pressure dependence match perfectly with the first mechanism in oxygen rich atmospheres, while the decrease of the energy of activation suggests a switch to the second path. Hence, the mechanism proposed in the previously mentioned work seems to be perfectly applicable in the case of interface-dominated strained LSM thin films (see figure 7). In air, the oxygen will be adsorbed and dissociated near the grain boundary and then a vacancy will diffuse reaching the atomic $O^-$ and finally incorporating it into the lattice. Lowering the $PO_2$ will originate a higher concentration of oxygen vacancies also on the grain bulk (although the grain boundaries remain the preferential path, as testified by the steady diffusion), causing a change in the *rds* and the consequential variation in the energy of activation. By assuming this mechanism, it is likely that the enhancement of $k^q$ measured is due to the high concentration of mobile vacancies present in the grain boundaries, which would act as a sink for the oxygen adsorbed on the surface. This picture, sketched in figure 7, can explain also the strong similarity in the activation energy observed in section 3.2 for $k^q$ and $D^q$ and the lowering of surface resistance decreasing the LSM grain size observed in [18]. Although we are aware that an exact calculation of the enhancement of the surface reaction rate would require the modelling of the grain boundary composition (oxygen vacancies and cations concentrations), an increase of k up to four orders of magnitude has been estimated from the grain boundary vacancy concentration extracted from the parameters of this work (see Supplementary information).

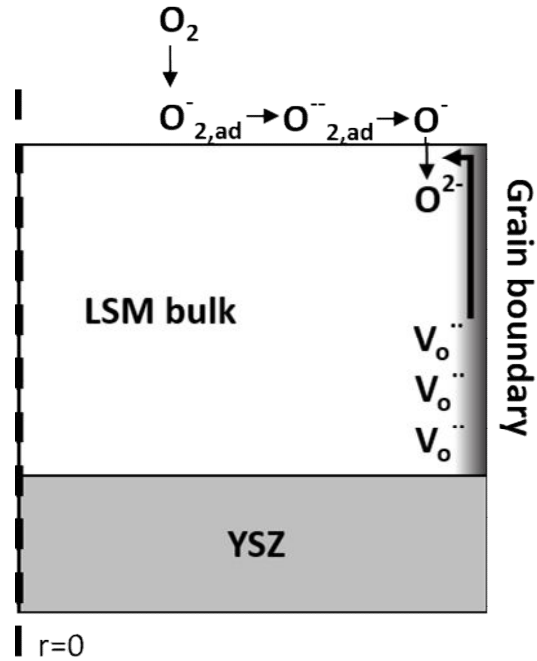

**Figure 7.** Sketch of the proposed oxygen reduction reaction mechanism taking place in air for interface-dominated strained LSM thin films. The coordinate $r=0$ refers to the center of the grain. The incorporation of the oxygen in the lattice takes place after oxygen vacancies coming from the grain boundary reach adsorbate species present at the surface. These adsorbate species come from dissociation of oxygen molecules without involving any oxygen vacancy.

**4. Conclusions**

Single phase dense columnar interface-dominated LSM thin films were deposited by PLD on single crystals of YSZ. LSM/YSZ/LSM symmetrical cells were characterized by EIS as a function of temperature and oxygen partial pressure to understand the dominating mechanisms in oxygen reduction reactions occurring when LSM acts as a cathode. A large enhancement of the oxygen diffusion with a lower dependence on oxygen partial pressure with respect to the bulk LSM is observed and discussed. A large concentration of defects in the grain boundaries is suggested to be in the origin of the major differences with bulk LSM, typically showing low oxygen diffusion and oxygen hyperstoichiometric in air. This high concentration of oxygen vacancies in the grain boundaries is expected to be related to a strong chemo-mechanical local field, which could not be modifiable by changes of $P_{O2}$ within the range covered in this work. A similar enhancement is presented for the oxygen surface exchange likely related to this large increase of oxygen vacancies in the grain boundaries, which makes the oxygen incorporation easier. An oxygen incorporation mechanism for the ORR is proposed. Accordingly, the grain boundary acts as a sink providing oxygen vacancies to the surface where the reduction of oxygen adsorbate species takes place. Nonetheless, the comprehension of the kinetics of LSM grain boundaries is still not fulfilled. Moreover, the large variability in the enhancement of the mass transport properties reported in the literature for similar LSM thin films suggests that further work is necessary to elucidate the origin and to define strategies to control and tune this superior performance.

**Acknowledgements**

The research was supported by the *Generalitat de Catalunya-AGAUR* (M2E exp. 2014 SGR 1638), the *Ministerio de Economía y Competitividad* (BELIF project, TEC2015-62519-ERC) and the European Regional Development Funds (ERDF, "FEDER Programa Competitivitat de Catalunya 2007-2013").

**References**

[1] S.B. Adler, Factors governing oxygen reduction in solid oxide fuel cell cathodes, Chem. Rev. 104 (2004) 4791–4843.

[2] R.A. De De Souza, J.A. Kilner, J.F. Walker, A SIMS study of oxygen tracer diffusion and surface exchange in LSM, Mater. Lett. 43 (2000) 43–52.

[3] J. Fleig, H.-R. Kim, J. Jamnik, J. Maier, Oxygen Reduction Kinetics of Lanthanum Manganite (LSM) Model Cathodes: Partial Pressure Dependence and Rate-Limiting Steps, Fuel Cells. 8 (2008) 330–337. doi:10.1002/fuce.200800025.

[4] F. Poulsen, Defect chemistry modelling of oxygen-stoichiometry, vacancy

concentrations, and conductivity of (La1−xSrx)yMnO3±δ, Solid State Ionics. 129 (2000) 145–162. doi:10.1016/S0167-2738(99)00322-7.

[5]  A. Hammouche, E. Siebert, A. Hammou, M. Kleitz, Electrocatalytic Properties and Nonstoichiometry of the High Temperature Air Electrode La[sub 1−x]Sr[sub x]MnO[sub 3], J. Electrochem. Soc. 138 (1991) 1212. doi:10.1149/1.2085761.

[6]  E. Siebert, A. Hammouche, M. Kleitz, Impedance spectroscopy analysis of La1-xSrxMnO3-yttria-stabilized zirconia electrode kinetics, Electrochim. Acta. 40 (1995) 1741–1753. doi:10.1016/0013-4686(94)00361-4.

[7]  S.B. Adler, Limitations of charge-transfer models for mixed-conducting oxygen electrodes, Solid State Ionics. 135 (2000) 603–612. doi:10.1016/S0167-2738(00)00423-9.

[8]  Y.A. Mastrikov, R. Merkle, E. Heifets, E.A. Kotomin, J. Maier, Pathways for oxygen incorporation in mixed conducting perovskites: A DFT-based Mechanistic analysis for (La, Sr)MnO3-??, J. Phys. Chem. C. 114 (2010) 3017–3027. doi:10.1021/jp909401g.

[9]  B.C.H. Steele, a Heinzel, Materials for fuel-cell technologies, Nature. 414 (2001) 345–352. doi:10.1038/35104620.

[10] K. Wen, W. Lv, W. He, Interfacial lattice-strain effects on improving the overall performance of micro-solid oxide fuel cells, J. Mater. Chem. A. 3 (2015) 20031–20050. doi:10.1039/C5TA03009A.

[11] J. Maier, Nanoionics: ionic charge carriers in small systems, Phys. Chem. Chem. Phys. 11 (2009) 3011–3022. doi:10.1039/b905911n.

[12] A.M. Saranya, D. Pla, A. Morata, A. Cavallaro, J. Canales-Vázquez, J.A. Kilner, M. Burriel, A. Tarancón, Engineering Mixed Ionic Electronic Conduction in La 0.8 Sr 0.2 MnO 3+ δ Nanostructures through Fast Grain Boundary Oxygen Diffusivity, Adv. Energy Mater. 5 (2015) n/a–n/a. doi:10.1002/aenm.201500377.

[13] E. Navickas, T.M. Huber, Y. Chen, W. Hetaba, G. Holzlechner, G. Rupp, M. Stöger-Pollach, G. Friedbacher, H. Hutter, B. Yildiz, J. Fleig, Fast oxygen exchange and diffusion kinetics of grain boundaries in Sr-doped LaMnO 3 thin films, Phys. Chem. Chem. Phys. 17 (2015) 7659–7669. doi:10.1039/C4CP05421K.

[14] F. Dion, A. Lasia, The use of regularization methods in the deconvolution of underlying distributions in electrochemical processes, J. Electroanal. Chem. 475 (1999) 28–37. doi:10.1016/S0022-0728(99)00334-4.

[15] H. Schichlein, A.C. Müller, M. Voigts, A. Krügel, E. Ivers-Tiffée, Deconvolution of electrochemical impedance spectra for the identification of electrode reaction mechanisms in solid oxide fuel cells, J. Appl. Electrochem. 32 (2002) 875–882. doi:10.1023/A:1020599525160.

[16] S. Hershkovitz, S. Baltianski, Y. Tsur, Harnessing evolutionary programming for impedance spectroscopy analysis: A case study of mixed ionic-electronic conductors, Solid State Ionics. 188 (2011) 104–109. doi:10.1016/j.ssi.2010.10.004.

[17] T.H. Wan, M. Saccoccio, C. Chen, F. Ciucci, Influence of the Discretization Methods on the Distribution of Relaxation Times Deconvolution: Implementing Radial Basis Functions with DRTtools, Electrochim. Acta. 184 (2015) 483–499.

doi:10.1016/j.electacta.2015.09.097.

[18] R.E. Usiskin, a S. Maruyama, b C.J. Kucharczyk, a I. Takeuchib, S.M. Haile, Probing the reaction pathway in (La0.8Sr0.2)0.95MnO3+δ using libraries of thin film microelectrodes, J. Mater. Chem. A. 3 (2015) 19330. doi:10.1039/c5ta02428e.

[19] Y. Zhang, Y. Chen, M. Yan, F. Chen, Reconstruction of relaxation time distribution from linear electrochemical impedance spectroscopy, J. Power Sources. 283 (2015) 464–477. doi:10.1016/j.jpowsour.2015.02.107.

[20] A. Häffelin, J. Joos, M. Ender, A. Weber, E. Ivers-Tiffée, Time-Dependent 3D Impedance Model of Mixed-Conducting Solid Oxide Fuel Cell Cathodes, J. Electrochem. Soc. 160 (2013) F867–F876. doi:10.1149/2.093308jes.

[21] J. Jamnik, J. Maier, Generalised equivalent circuits for mass and charge transport: chemical capacitance and its implications, Phys. Chem. Chem. Phys. 3 (2001) 1668–1678. doi:10.1039/b100180i.

[22] J. Bisquert, Theory of the impedance of electron diffusion and recombination in a thin layer, J. Phys. Chem. B. 106 (2002) 325–333. doi:10.1021/jp011941g.

[23] J. Maier, On the correlation of macroscopic and microscopic rate constants in solid state chemistry, Solid State Ionics. 112 (1998) 197–228. doi:10.1016/S0167-2738(98)00152-0.

[24] T. Ishigaki, S. Yamauchi, K. Kishio, J. Mizusaki, K. Fueki, Diffusion of oxide ion vacancies in perovskite-type oxides, J. Solid State Chem. 73 (1988) 179–187. doi:10.1016/0022-4596(88)90067-9.

[25] W.C. Chueh, W. Lai, S.M. Haile, Electrochemical behavior of ceria with selected metal electrodes, Solid State Ionics. 179 (2008) 1036–1041. doi:10.1016/j.ssi.2007.12.087.

[26] R.A. De Souza, J. a. Kilner, Oxygen transport in La1−xSrxMn1−yCoyO3±δ perovskites: Part II. Oxygen surface exchange, Solid State Ionics. 126 (1999) 153–161.

[27] T.M. Huber, M. Kubicek, a. K. Opitz, J. Fleig, The Relevance of Different Oxygen Reduction Pathways of La0.8Sr0.2MnO3 (LSM) Thin Film Model Electrodes, J. Electrochem. Soc. 162 (2015) F229–F242. doi:10.1149/2.0061503jes.

[28] K. Develos-Bagarinao, H. Kishimoto, J. De Vero, K. Yamaji, T. Horita, Effect of La0.6Sr0.4Co0.2Fe0.8O3-δ microstructure on oxygen surface exchange kinetics, Solid State Ionics. 288 (2016) 6–9. doi:10.1016/j.ssi.2016.01.008.

[29] L. Yan, P. a Salvador, Substrate and Thickness Effects on the Oxygen Surface Exchange of La0.7Sr0.3MnO3 Thin Films, ACS Appl. Mater. Interfaces. 4 (2012) 2541–2550. doi:10.1021/am300194n.

[30] R. Das, D. Mebane, E. Koep, M. Liu, Modeling of patterned mixed-conducting electrodes and the importance of sheet resistance at small feature sizes, Solid State Ionics. 178 (2007) 249–252. doi:10.1016/j.ssi.2006.12.021.

[31] T. Ishihara, ed., Perovskite Oxide for Solid Oxide Fuel Cells, Springer US, Boston, MA, 2009. doi:10.1007/978-0-387-77708-5.

[32] T.M. Huber, E. Navickas, G. Friedbacher, H. Hutter, J. Fleig, Apparent Oxygen Uphill Diffusion in La 0.8 Sr 0.2 MnO 3 Thin Films upon Cathodic

Polarization, ChemElectroChem. 2 (2015) 1487–1494. doi:10.1002/celc.201500167.

[33] D. Chen, S.R. Bishop, H.L. Tuller, Non-stoichiometry in Oxide Thin Films: A Chemical Capacitance Study of the Praseodymium-Cerium Oxide System, Adv. Funct. Mater. 23 (2013) 2168–2174. doi:10.1002/adfm.201202104.

[34] J. Nowotny, M. Rekas, Defect Chemistry of (La,Sr)MnO3, J. Am. Ceram. Soc. 81 (1998) 67–80. doi:10.1111/j.1151-2916.1998.tb02297.x.

[35] L. Wang, R. Merkle, Y. a. Mastrikov, E. a. Kotomin, J. Maier, Oxygen exchange kinetics on solid oxide fuel cell cathode materials—general trends and their mechanistic interpretation, J. Mater. Res. 27 (2012) 2000–2008. doi:10.1557/jmr.2012.186.

[36] R. a De Souza, A universal empirical expression for the isotope surface exchange coefficients ($k^*$) of acceptor-doped perovskite and fluorite oxides., Phys. Chem. Chem. Phys. 8 (2006) 890–897. doi:10.1039/b511702j.

[37] W. Jung, H.L. Tuller, A New Model Describing Solid Oxide Fuel Cell Cathode Kinetics: Model Thin Film $SrTi_{1-x}Fe_xO_{3-\delta}$ Mixed Conducting Oxides–a Case Study, Adv. Energy Mater. 1 (2011) 1184–1191. doi:10.1002/aenm.201100164.

[38] J. Fleig, R. Merkle, J. Maier, The $p(O_2)$ dependence of oxygen surface coverage and exchange current density of mixed conducting oxide electrodes: model considerations., Phys. Chem. Chem. Phys. 9 (2007) 2713–2723. doi:10.1039/b618765j.

[39] Y.A. Mastrikov, R. Merkle, E. Heifets, E.A. Kotomin, J. Maier, Pathways for oxygen incorporation in mixed conducting perovskites: A DFT-based Mechanistic analysis for (La, Sr)$MnO_{3-\delta}$, J. Phys. Chem. C. 114 (2010) 3017–3027. doi:10.1021/jp909401g.

# Supplementary information

**Insights into the enhancement of oxygen mass transport properties of strontium-doped lanthanum manganite interface-dominated thin films**

F. Chiabrera[1], A. Morata[1], A. Tarancón[1,*]

[1]*Department of Advanced Materials for Energy, Catalonia Institute for Energy Research (IREC), Jardins de les Dones de Negre 1, Planta 2, E-08930 Sant Adrià de Besòs (Barcelona), Spain, e-mail: atarancon@irec.cat*

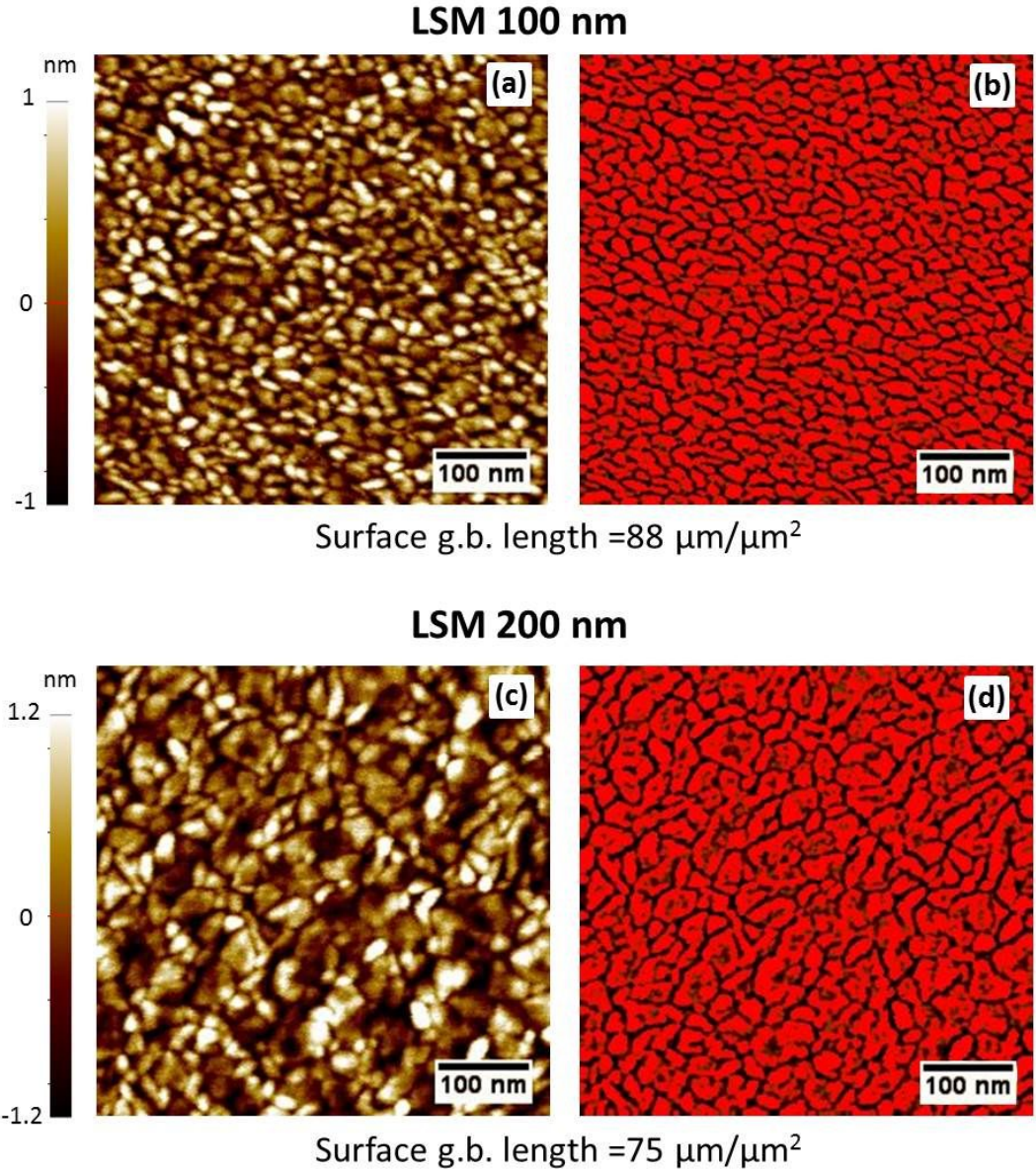

**Figure S1.** AFM images of the (a) 100 nm sample and (c) 200 nm one as taken and after image processing used to calculate the exposed grain boundary length per unit area.

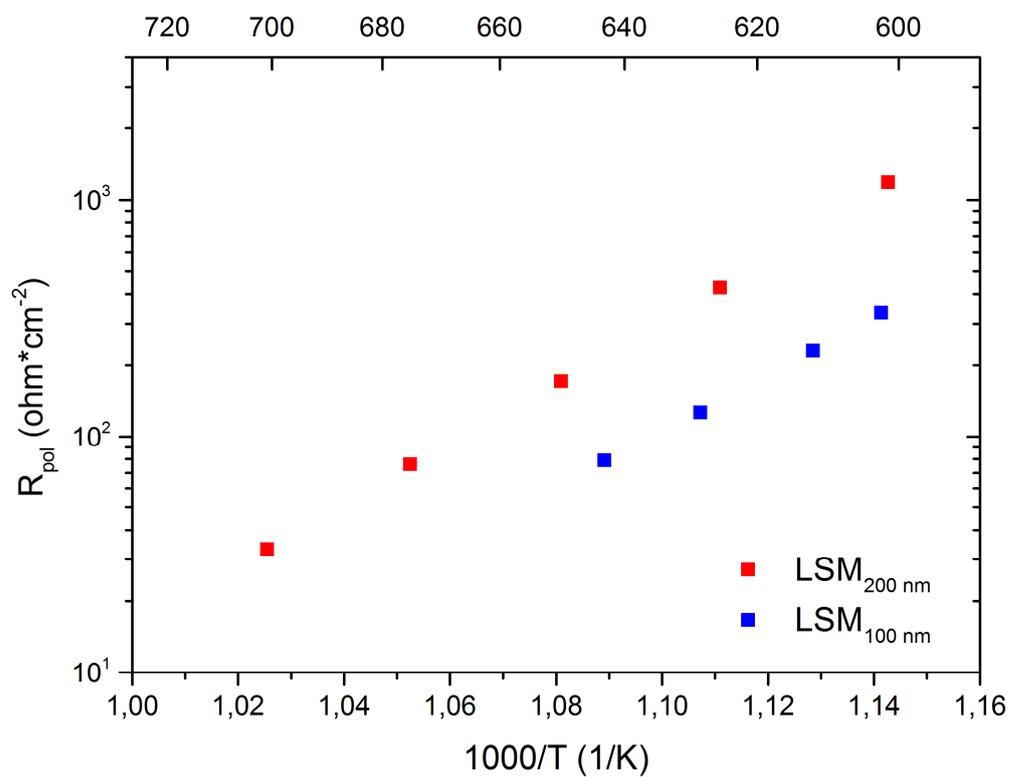

**Figure S2.** Total polarization resistance calculated from the 100 nm and 200 nm LSM thin film at different temperature values.

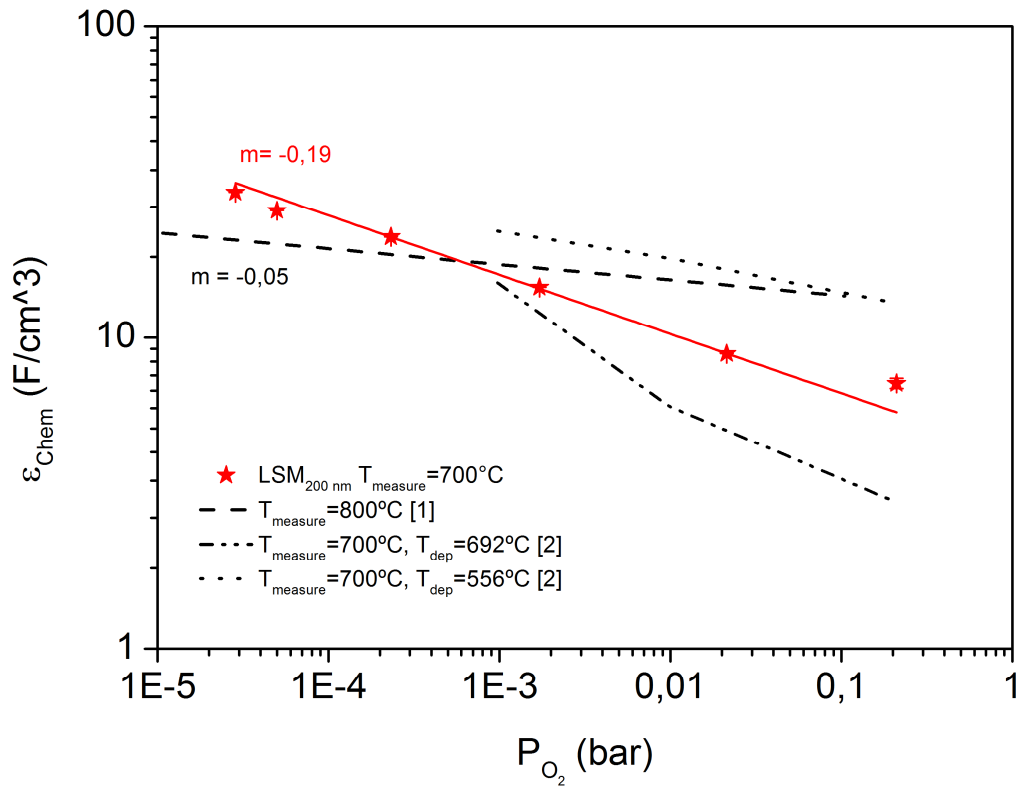

**Figure S3.** Partial pressure dependence of the volume normalized chemical capacitance for the 200nm-thick electrodes LSM/YSZ/LSM sample at 700ºC. Value extracted from [1] (dashed line) and [2] (short dashed line: high grain boundary density; dash dot dotted line low grain boundary density) are included for comparison.

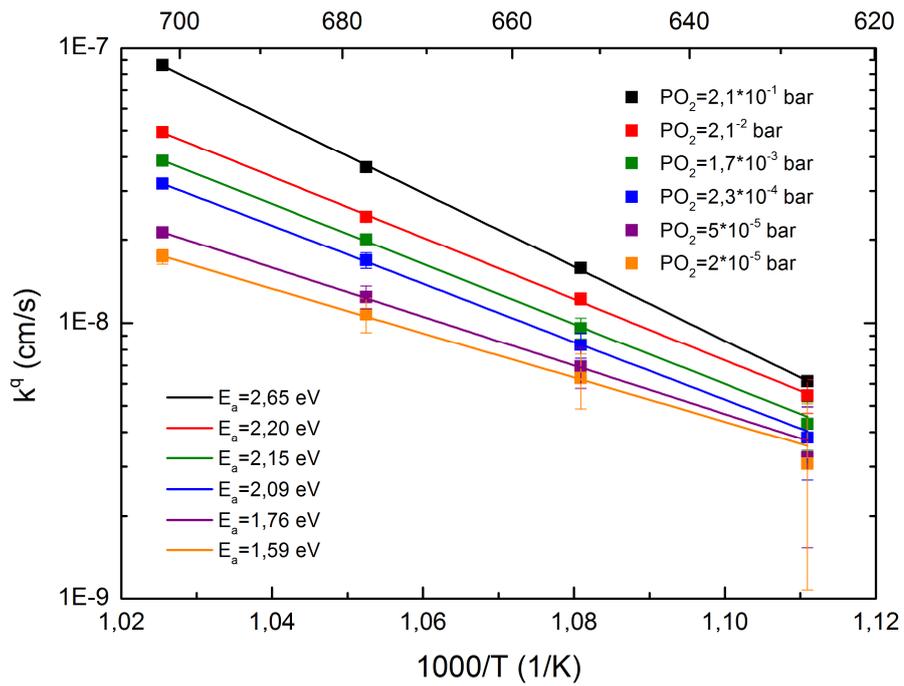

**Figure S4.** Temperature dependence of the surface exchange coefficient for the 200nm-thick electrode LSM/YSZ/LSM sample at different oxygen partial pressure. The activation energy of each $P_{O2}$ is shown in the lower left part of the figure.

# Estimation of the surface reaction rate enhancement

Since for LSM the most probable *rds* in the oxygen reduction reactions is the approach of an oxygen vacancy to an adsorbed O$^-$ (as found by Maier and co-workers in [3]), the reaction rate $\mathcal{R}$ is proportional to $\mathcal{R} \propto k^q \propto c_v \cdot D_v \cdot \theta(O^-)$, where $c_v$ is the concentration of vacancies, $D_v$ is the vacancy diffusion and $\theta(O^-)$ is the adsorbate coverage. It is possible to estimate the mole fraction of vacancy in the grain boundaries as $x_{v,gb} = D^q_{gb}/D^\delta_{gb}$, with $D^\delta_{gb}$ being the grain boundary chemical diffusion (calculated as $\sigma_{ion,gb}/\varepsilon_{chem,gb}$) and considering the grain boundary length per unit area extracted from the AFM analysis (Fig. S1) and a grian boundary width of $\delta_{gb}=1$ nm. The extracted value for the 200 nm sample at 700ºC in air is $c_{v,gb} = x_{v,gb} \cdot c_{o2-} = 2 \cdot 10^{20}$ cm$^{-3}$, over 6 order of magnitude higher than the bulk value measured for LSM [4]. It is important to notice that this calculation does not take in account space charge effects and compositional homogeneities that are supposed to take place in the LSM, therefore it can be considered just as a first approximation. It is known that a higher concentration of vacancy in the LSM first atomic layer is expected due to a lower formation reaction energy, with a calculated concentration of $c_{v,s} = 6 \cdot 10^{19}$ cm$^{-3}$ [3]. Nevertheless, the surface vacancies are expected to be present just in the very first atomic layer, while, due to the high $D_v$ (measured in bulk LSM to be ~$10^{-7}$cm$^2$/s at 700ºC [5]), in the grain boundary also the vacancies coming from the film could partecipate to the reaction. On these basis, one can calculate the highest enhancment to the surface rate reaction considering that all the vacancies in the grain boundary can partecipate to the reaction, as $k_{gb}/k_{bulk} \propto (c_{v,gb} \cdot \delta_{gb} \cdot p_{gb} \cdot A \cdot d)/(c_{v,surf} \cdot \delta_s \cdot A)$, where $\delta_{gb}$ is the grain boundary length, $p_{gb}$ is the surface grain boundary length, A the surface area and d the thickness of film. A reaction rate increase up to 4 order of magnitude is founded (depending on the active thickness), which seems to justify the enhancemnt systematically found. As mentioned before, this calculation is just a first estimation, since possible compositional changes could take place in the grain boundary (space charge layer) leading to differences in the reaction rate.

A final interesting observation is that the two mechanism of surface and grain boundary diffusion could both contribute to the oxygen incorporation, determining a complex 3D interaction.

**References**

[1] J. Fleig, H.-R. Kim, J. Jamnik, J. Maier, Oxygen Reduction Kinetics of Lanthanum Manganite (LSM) Model Cathodes: Partial Pressure Dependence and Rate-Limiting Steps, Fuel Cells. 8 (2008) 330–337. doi:10.1002/fuce.200800025.

[2] R.E. Usiskin, a S. Maruyama, b C.J. Kucharczyk, a I. Takeuchib, S.M. Haile, Probing the reaction pathway in (La0.8Sr0.2)0.95MnO3+δ using libraries of thin film microelectrodes, J. Mater. Chem. A. 3 (2015) 19330. doi:10.1039/c5ta02428e.

[3] Y.A. Mastrikov, R. Merkle, E. Heifets, E.A. Kotomin, J. Maier, Pathways for oxygen incorporation in mixed conducting perovskites: A DFT-based Mechanistic analysis for (La, Sr)MnO3-δ, J. Phys. Chem. C. 114 (2010) 3017–3027. doi:10.1021/jp909401g.

[4] I. Yasuda, K. Ogasawara, M. Hishinuma, T. Kawada, M. Dokiya, Oxygen tracer diffusion coefficient of (La,Sr)MnO3+/-delta, Solid State Ionics. 86-8 (1996) 1197–1201.

[5] R. De Souza, J. a. Kilner, Oxygen transport in La1−xSrxMn1−yCoyO3±δ perovskites Part I. Oxygen tracer diffusion, Solid State Ionics. 106 (1998) 175–187. doi:10.1016/S0167-2738(97)00499-2.